\documentclass[aps,prl,nofootinbib,,superscriptaddress,groupedaddress, preprintnumbers,twocolumn]{revtex4}
\usepackage{amsmath}
\usepackage{amsfonts}
\usepackage{amssymb}
\usepackage{dsfont}
\usepackage{mathrsfs}
\usepackage{graphicx}
\usepackage{epstopdf}
\usepackage{color}
\usepackage[all]{xy}
\usepackage{hyperref}
\hypersetup{
    colorlinks=true, 
    linktoc=page,    
    linkcolor=blue,  
    urlcolor=black
   }
\newcommand{\ie}{\emph{i.e.~}}
\newcommand{\eg}{\emph{e.g.~}}

\newcommand{\hphi}{\hat{\varphi}}
\newcommand{\hphid}{\hat{\varphi}^{\dagger}}

\newcommand{\hsigma}{\hat{\sigma}}
\newcommand{\hsigmad}{\hat{\sigma}^{\dagger}}

\newcommand{\move}{\widehat{\mathcal{M}}}

\newcommand{\ket}[1]{\left| #1 \right \rangle}

\newcommand{\extd}{\mathrm{d}}

\newcommand{\SU}{\mathrm{SU}}

\newcommand{\va}{\scriptscriptstyle}

\newcommand{\be}{
\begin{equation}}
\newcommand{\ee}{\end{equation}}

\newcommand{\bee}{\nopagebreak[3]\begin{equation*}}
\newcommand{\eee}{\end{equation*}}

\newcommand{\ba}{\nopagebreak[3]\begin{eqnarray}}
\newcommand{\ea}{\end{eqnarray}}

\newcommand{\baa}{\nopagebreak[3]\begin{eqnarray*}}
\newcommand{\eaa}{\end{eqnarray*}}

\newcommand{\la}{\label}
\newcommand{\n}{\nonumber}

\newcommand{\C}{\mathbb{C}}
\newcommand{\N}{\mathbb{N}}

\newcommand{\iu}{\mathrm{i}}

\newcommand{\AIH}{{\mathcal A}_{\va H}}

\usepackage[dvipsnames]{xcolor}

\begin{document}

\title{Horizon Entropy from Quantum Gravity Condensates}
\author{Daniele Oriti}
\email{daniele.oriti@aei.mpg.de}
\affiliation{Max Planck Institute for Gravitational Physics (AEI), 
Am M\"uhlenberg 1, D-14476 Golm, Germany}
\author{Daniele Pranzetti}
\email{dpranzetti@sissa.it}
\affiliation{Scuola Internazionale Superiore di Studi Avanzati (SISSA), via Bonomea 265, 34136 Trieste, Italy}
\author{Lorenzo Sindoni}
\email{lorenzo.sindoni@aei.mpg.de}
\affiliation{Max Planck Institute for Gravitational Physics (AEI), 
Am M\"uhlenberg 1, D-14476 Golm, Germany}

\begin{abstract}
We construct condensate states encoding the continuum spherically symmetric quantum geometry of a horizon in full quantum gravity, i.e. without any classical symmetry reduction, in the group field theory formalism. Tracing over the bulk degrees of freedom, we show how the resulting reduced density matrix manifestly exhibits a holographic behavior. We derive a complete orthonormal basis of eigenstates for the reduced density matrix of the horizon and use it to compute the horizon entanglement entropy. By imposing consistency with the horizon boundary conditions and semiclassical thermodynamical properties, we recover the Bekenstein--Hawking entropy formula for any value of the Immirzi parameter. Our analysis supports the equivalence between the von Neumann (entanglement) entropy interpretation and the Boltzmann (statistical) one.

 \end{abstract}

\preprint{}

\maketitle





{\bf Introduction}.---In this Letter 
we build and analyze, for the first time,
spherically symmetric continuum states to model a quantum black hole horizon, working
in the full theory. In doing so, we make no reference to a classical symmetry-reduced sector \cite{IH-rev}. 

As quantum gravity states for continuum spherically symmetric geometries we use 
spin network condensates in the group field theory (GFT) formalism \cite{GFT,OPRS},
 a framework in which the concepts and tools of loop quantum gravity (LQG) and spinfoam models are implemented and extended in a second quantized language. 
We impose on them conditions characterizing horizons, 
 and analyze their entanglement properties. 
We show that their entanglement entropy coincides with the Boltzmann entropy of horizon degrees of freedom (DOF) and it satisfies an area law, 
a cornerstone of holography.

The major strength of our analysis is the possibility of keeping into account the sum over triangulations required in the coarse graining procedure 
leading 
to an effective macroscopic description, 
as well as the control over the interplay between horizon boundary conditions and the calculation of entropy.
In fact, we are able to control the states with a relatively small number of parameters, encoding the geometrical data of the continuum geometry: 
we are using hydrodynamic states.
This construction allows us to explicitly 
compute the horizon 
density matrix
and to prove the holographic nature of our states. 

The implications of these novel features are striking. The entanglement entropy can be
computed exactly and it matches the Bekenstein--Hawking formula \cite{BH} for {\it any value} of the Immirzi parameter $\gamma$ (see \cite{gamma, KMS, gamma2} for a  discussion on
$\gamma$), once consistency with semiclassical conditions is imposed. 
The calculation reduces to a state counting, with the microscopic DOF encoded in the combinatorial structure of all possible horizon condensate graphs (for a fixed  macroscopic area value). This supports the equivalence between
entanglement  and statistical
interpretations of black hole entropy suggested in \cite{KMS}.

Because of difficulties in extracting effective equations of motion for 
our generalized GFT condensates 
from the fundamental dynamics of a given GFT model, we will omit restrictions originating from the microscopic dynamics in this work.
However, we will rely on the use a maximum entropy principle to capture a few essential
dynamical features, and as a partial characterization of horizon geometries, and we will show the consequence of requiring the compatibility with the classical 
dynamics of horizons and their thermodynamical properties.
\vspace{0.07cm}

{\bf Construction}.---Our plan consists of the following steps. 
$(i)$ We define GFT condensate states (as constructed in \cite{OPRS}) for a  
spacelike, 
spherically symmetric geometry by acting with a class of refinement operators on a seed state, and with appropriate semiclassicality restrictions. $(ii)$ We derive the reduced density matrix, tracing away the remaining bulk DOF and find a complete orthonormal basis of its eigenstates. $(iii)$ We compute the entanglement entropy, coinciding with the statistical entropy of the boundary DOF, and show how the result is affected by different choices of 
boundary conditions. 
\vspace{0.07cm}

{\it Spherically symmetric quantum states.}---\label{sec:BH}We define a spherical symmetric quantum geometry in terms of a gluing of homogeneous spherical shells to one another \cite{OPRS}. 
The  
states of each 
shell are constructed starting from a seed state for a given shell, upon which we act with refinement operators, increasing the number of  vertices and keeping the topology fixed as the connectivity is changed. 
In this way, the GFT state for a given shell is given by a (possibly infinite) superposition of regular 4-valent graphs with given topology. Shells are then glued together to form a full 3D foliation.

To keep the topological
structure under control, each 4-vertex carries a color $t=\{B,W\}$ and each $\SU(2)$ group element $g$ associated to a link of a given 4-vertex is labelled by a number $I=\{1,2,3,4\}$ (i.e. we use colored 4-graphs \cite{tensor}). Each shell is composed of three parts: an outer boundary, an inner boundary, and a bulk in between. In order to distinguish these regions, we introduce a further color $s=\{+,0,-\}$, specifying whether a given vertex belongs to the outer boundary, to the bulk or to the inner boundary, respectively.
The initial seed state and the refinement operators are such that all the open radial links of each boundary have the same color, different for the two boundaries. 
{In order to glue shells together, and still be able to distinguish different shells, we 
add a label $r\in \N$ 
to the shell wave-function, which effectively plays the role of a radial coordinate. } 
Two shells $r$ and $r+1$ are then
   glued together 
   through their radial links as schematically depicted here
 \begin{figure}[h]
   \centerline{\(
   \begin{array}{c}
   \includegraphics[height=2.5cm]{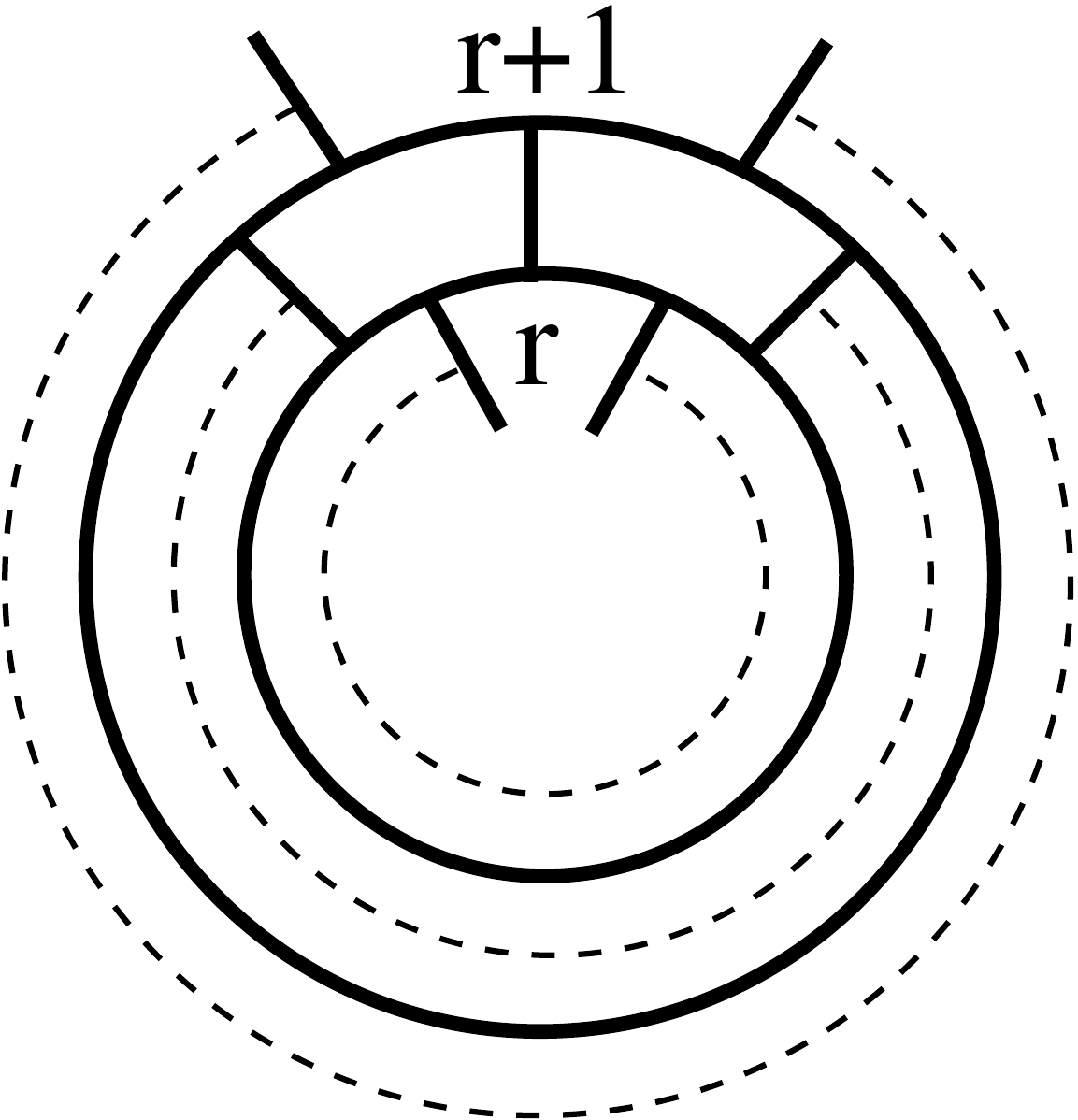}\,.
   \end{array}\)}
 \end{figure}
 
The idea of GFT condensation posits that the same wave-function $\sigma$ should be associated to each 
new GFT excitation introduced in the state. This notion of wave-function homogeneity {\it for each shell} captures the coarse grained homogeneity of continuum geometric data \cite{OPRS}. 
Our construction relies on the operatorial version of GFT, which provides a second quantization formalism for LQG \cite{GFT}. The main advantage of this formulation is that it allows us to introduce a Fock space structure in the description of spin network states. More precisely, the field ladder operators for the vertex $v$ are constructed from the original GFT field creation/annihilation operators 
 satisfying the  bosonic commutation relations:
\begin{equation}
	\label{eq:BosComRel}
	[\hphi_ {t^{\va v}}(g^v_I), \hphid_ {t^{\va w}}(g^w_I)] 
	=\int_{\SU(2)} d\gamma 
	 \prod_{I=1}^4 \delta(g^v_I \gamma (g^w_I)^{-1})
\end{equation}
and having a graph-theoretic interpretation as operators creating or destroying 4-valent vertices.
From these, we can then define field operators encoding the wave-function associated to the vertices and  incorporating the vertex homogeneity idea. More precisely, we introduce the transformed fields 
\ba
\hsigma_{r,t^{\va v}s^{\va v}}(h^v_I) = \int \extd g^v_I\; \sigma_{r,s^{\va v}}(h^v_Ig^v_I) \,\hphi_ {t^{\va v}}(g^v_I)
\la{c-field}
\ea
and its adjoint,  satisfying the commutation relations
\be\la{c-comm}
\left[\hsigma_{r,t^{\va v}s^{\va v}}(h^v_I), \hsigmad_{r',t^{\va w}s^{\va w}}(h^w_I) \right]  =
\delta_{r,r'}\delta_{t^{\va v}, t^{\va w}}\delta_{s^{\va v}, s^{\va w}}\Delta_{L}(h^v_I, h^w_I). 
\ee
Here we have defined the left invariant Dirac delta, appearing due to gauge invariance properties of the operators wave-function, as:
$
\Delta_{L}(h^v_I, h^w_I)=\int_{SU(2)} d\gamma \prod_{I=1}^4 \delta(\gamma h^v_I (h^w_I)^{-1})
$.
The choice of the factor $\delta_{r,r'}$ in the commutator is crucial: it implies that operators associated to different shells commute with each other. The commutator \eqref{c-comm} was introduced in \cite{OPRS} for 
technical reasons, but we will show that it encodes crucial physical properties, as the form of \eqref{c-comm} is at the origin of the holographic nature of our states.

All the non-radial links departing from the 4-valent vertices  forming a given shell are glued among each other, through the convolution of group field arguments, so to construct 4-regular graphs associated to the Fock space states. This gluing  is mirrored in the dual by the gluing of tetrahedra to construct a three-dimensional simplicial shell topology.
A full spatial foliation can then be formed by glueing all the radial links of the outer boundary of the shell $r$ with the (same number of) radial links of the inner boundary of the shell $r+1$. Both sets of links must have the same color. 
We are not going to explicitly define a 
refinement operator for the glued shells,
as it plays no role in our entropy calculations (but see \cite{OPRS} for the tools used in the construction).
{The general expression for the full  
states that we are interested in, then, is of the type:} 
\begin{equation}\la{full-state}
\ket{\Psi} = \prod_{r} f_{r}(\move_{r,\va B},\move_{r,\va W}) \ket{seed}\,,
\end{equation}
where $f_r$ is a function of the refinement operators $\move_{r,t}$ of a given shell $r$. 
The seed state is constructed out of the graph with the minimal number of vertices necessary to encode the desired shell topology (the explicit form is given in \cite{OPRS}).
For a given shell $r$, the operators  $\move_{r,t}$ comprise the sum of three terms, namely $\move_{r,t}=\sum_{s=\{+,0,-\}}\move_{r,ts}$, each refining separately one part of the shell. These are examples of closed graph-labeled operators, constructed out of simple convolutions of three creation and one annihilation operators encoding the triangulation of a ball. For instance, in the case of $t=B, s=+,-$ with the arbitrary choice of radial links having color 1 (to which we associate the identity group element $h_1=e$), we have
\begin{align*}
\move_{r,{\va B}s}
\equiv \!\int 
 (dh)^6 
&\hsigma^{\dagger}_{r,\va B-}(e,h_{2},h_{3},h_{4'})
\hsigma^{\dagger}_{r,\va W-}(e,h_{2'},h_{3'},h_{4'})\\
&\hsigma^{\dagger}_{r,\va B-}(e,h_{2'},h_{3'},h_{4})
\hsigma_{r,\va B-}(e,h_{2},h_{3},h_{4})\,.
\end{align*}
The action of the refinement operators is defined through the commutator $[\move_{r,ts}, \hsigmad_{r,ts}(h_I)]$, corresponding to a dipole insertion in the given component $s$ of the shell $r$ and it can be represented pictorially as
\ba\la{MB}
&&\move_{r,{\va B}s}:~~
\begin{array}{c}
\includegraphics[width=1.0cm]{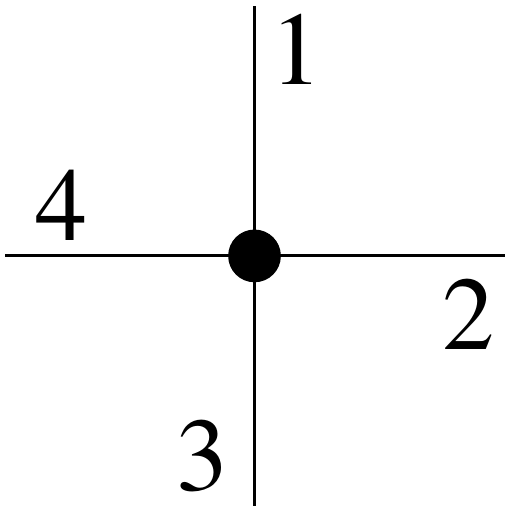}
\end{array}~~~\rightarrow~~~
\begin{array}{c}
\includegraphics[width=3.0cm]{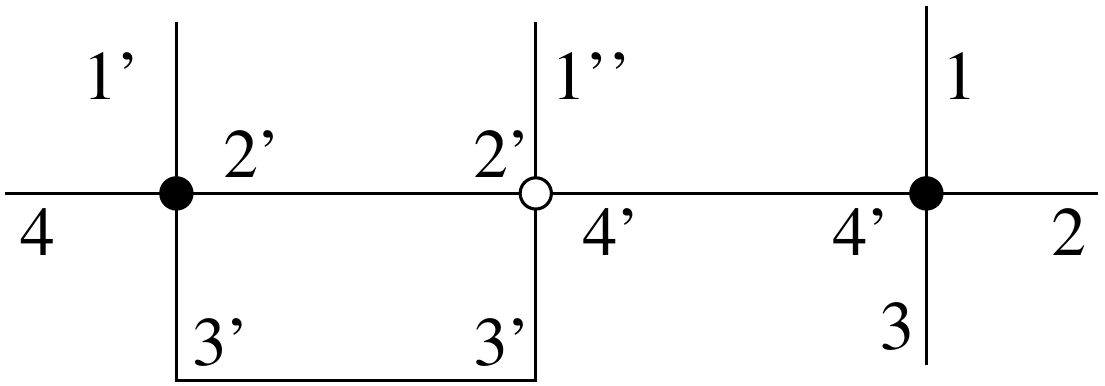}
\end{array}\\
&&\move_{r,{\va W}s}:~~
\begin{array}{c}
\includegraphics[width=1.0cm]{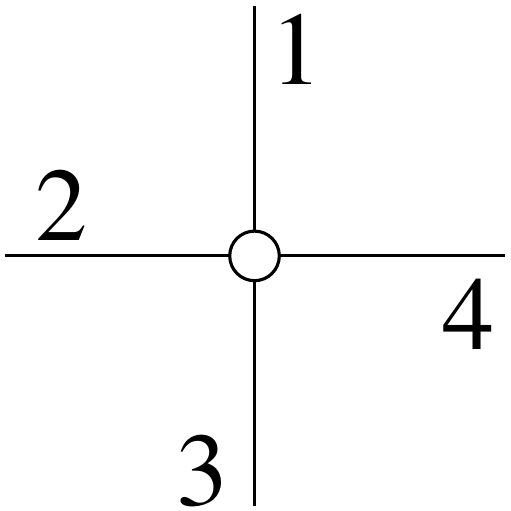}
\end{array}~~~\rightarrow~~~
\begin{array}{c}
\includegraphics[width=3.0cm]{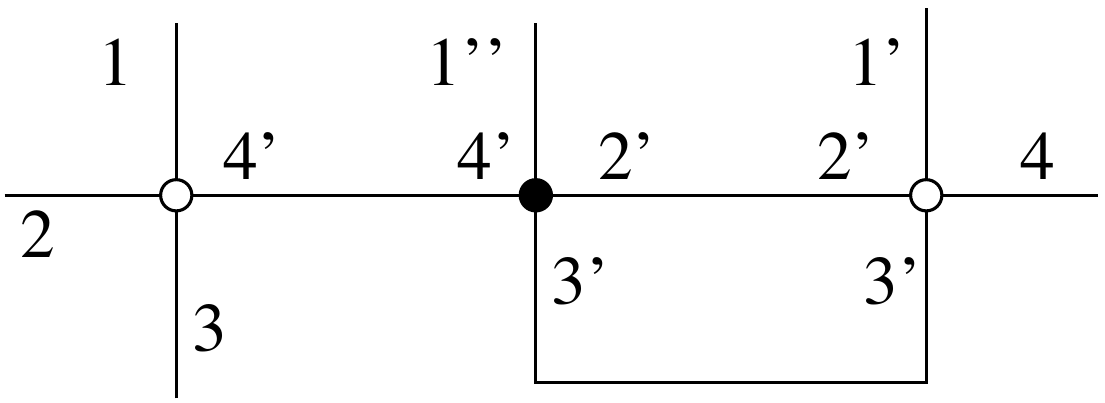}\la{MW}\,.
\end{array}
\ea

The form of the function $f_r$ amounts to fixing the coefficients of the linear combination of graphs, i.e. components with a fixed number of particles, appearing in the decomposition of the full shell state. This choice does not affect the leading term in the entropy calculation performed below, where we make one that keeps to a minimum the additional parameters controlling the state, but only the subleading logarithmic corrections, as we point out in the concluding remarks.

Geometric operators can then be computed for our GFT states, in a 2nd quantized language.
For example, following \cite{OPRS}, we define the horizon area operator 
\begin{equation}\la{area}
\hat{\mathbb{A}}_{Jr,s}
\equiv\sum_{t=\va {B,W}} \int d h^v_I \hsigmad_{r,ts}(h^v_I) \sqrt{E^{i}_{J} E^{j}_{J} \delta_{ij}} 
\hsigma_{r,ts}(h^v_I)\,,
\end{equation}
where in this case $s=\{+,-\}$ and $J$ corresponds to the color of the radial links dual to the boundary $s$ of the shell $r$ under examination. 
 The action of the operator \eqref{area} is computed using the definition
\begin{equation}
E^{i}_{J} \rhd f(g_I) := \lim_{\epsilon\rightarrow 0} \iu \frac{d}{d\epsilon}
f(g_{1},\ldots, e^{-\iu \epsilon \tau^{i}}g_{J}, \ldots, g_{4})
\end{equation}
for a given function $f:SU(2)^4\rightarrow \C$.
The expectation value of the area operator \eqref{area} on a shell boundary state 
gives
\begin{equation}\la{expectation}
\langle \hat{\mathbb{A}}_{Jr,s}  \rangle 
\, = \, \langle \widehat{n}_{r,s}  \rangle \, a_{Jr,s} ,
\end{equation}
where  
\be\la{a}
a_{Jr,s}=\int dh^v_I dg^v_I 
\sigma_{r,s}(h^v_I g^v_I)  \sqrt{E^{i}_{J} E^{j}_{J} \delta_{ij}} \rhd \overline{\sigma_{r,s}(h^v_I g^v_I)}
\ee
is the expectation value of the first quantized (LQG) area operator of a single radial link-$J$, in the boundary $s$ of the shell $r$, in a single-vertex state with wave-function $\sigma$; $\widehat{n}_{r,s}$ is the number operator defined as
$
\widehat{n}_{r,s}
=\sum_{t=\va {B,W}} \int dh_I\, \hsigmad_{r,ts} (h_I) \hsigma_{r,ts} (h_I)
$.
Because of the definition of the states, at each stage of refinement we always have 
$
n_{r, {\va B} s}=n_{r,{\va W} s}={n_{r,s}}/{2}\,,
$
where $n\equiv\langle\widehat{n}\rangle $.

{Notice that, in general, these expressions require regularization, as
our condensate states are not always normalizable 
\cite{OPRS}.
 However, it is easy to construct condensate states, peaked in some spin representation, for which all these steps can be followed rigorously. 
 The full space of 
solutions to the equations characterizing the condensate wave-function and the refinement move kernel is not known, and we can only exhibit a few explicit solutions. The existence of several other solutions is plausible, which then leaves
a certain amount of freedom in the specification of the vertex wave-function.

Let us point out that the existence of a number operator in the GFT formulation of LQG 
represents a key difference with respect to the standard formulation, and it has a crucial role in 
the entropy calculation below.
\vspace{0.07cm}


{\it Further restrictions.}---In this context, we have two possible ways to characterize our shell condensate  as a quantum horizon. One possibility would be to impose the quantum version of the classical  isolated horizons (IH) boundary condition \cite{IH-2}. This can be done locally, at the level of each single-vertex, by relating the curvature around the link dual to the boundary face to the flux associated with it, leading to a restriction on the vertex wave-function. A second way to define the horizon shell is 
through 
the condition that the 
reduced states maximize the entropy. 
Imposition of these two constraints in general does not  commute, and will give different results for the entropy. We come back on the strategy we follow after deriving the general result. 

Further restrictions on our states come from semiclassicality conditions: the fluctuations of a set of operators, \eg the area, should be small.
They restrict the possible superposition of graphs with a different number of vertices, as it is evident from \eqref{expectation}. Furthermore, we have to impose that the shells are thin, for the geometry to look smooth. This imposes
a restriction on the expectation value of the volume per shell, the transverse area and the number of nodes. We do not discuss explicitly the 
operator equation counterparts of these conditions, as they do not enter directly in our entropy calculations.
\vspace{0.07cm}

{\it Reduced density matrix}.---Now we focus on the computation of the entropy associated to the quantum horizon, as defined by our states. We do this in two steps: reduction to the density matrix associated with the outmost shell, and explicit computation of its entropy.
Our complete quantum state, described by the pure density matrix $
\hat \rho=|\Psi\rangle\langle \Psi|\,
$, consists of a (thin) shell and bulk DOF. 
We need only the DOF 
of  the outer boundary of the horizon shell $r_0$,
described by a reduced density matrix. 

A simple case will clarify the general procedure.
Consider the graph $A$ for the horizon outer boundary $r_0$ and the graph $B$ of the inner boundary of the neighboring shell $r_0+1$, glued along boundaries of color $1$ (while tangent links departing from vertices in $A$ are glued among each others, and similarly for $B$). In order to be properly glued they must have the same number of vertices $n$. 
The wave-function is 
\baa
&&\psi(g^{{\va A_1}},...,g^{{\va A_n}},g^{{\va B_1}},...,g^{{\va B_n}})
= \int \prod_{i=1}^{n}
 dh^{\va A_i}_I
 dh^{\va B_i}_I \n\\
&&\times\,
\sigma_{\va A_i}(h^{\va A_i}_I g^{\va A_i}_I )
\sigma_{\va B_i}(h^{\va B_i}_I g^{\va B_i}_I )
\prod_{J=1}^4 \delta(h^{\va v_i}_{J}(h^{t_{J^{\va v_i}}}_J)^{-1})\,,
\eaa
where the product of $\delta$'s encodes the connectivity of the total graph $A\cup B$. 
{The notation is designed to keep
track of the combinatorics in terms of vertices $v_i$ (with $v=A, B$) and edges $J$ of the graph, so that
$t_{J^{\va v_i}}$ is the target vertex of the edge $J$ departing from the vertex $v_i$}. 
The total density matrix is
\baa
&&\rho^{(n)}
=\int \prod_{i=1}^{n}
dh^{\va A_i}_I
 dh^{\va B_i}_I
dk^{\va A_i}_I
dk^{\va B_i}_I\\
&&\times\,\left(\sigma_{\va A_i}(h^{\va A_i}_I g^{\va A_i}_I )
\sigma_{\va B_i}(h^{\va B_i}_I g^{\va B_i}_I )
\prod_{J=1}^4 \delta(h^{\va v_i}_{J}(h^{t_{J^{\va v_i}}}_J)^{-1})\right)\\
&&\times\,\left(\overline{\sigma_{\va A_i}(k^{\va A_i}_I g'^{\va A_i}_I )
\sigma_{\va B_i}(k^{\va B_i}_I g'^{\va B_i}_I )}
\prod_{J=1}^4 \delta(k^{\va v_i}_{J}(k^{t_{J^{\va v_i}}}_J)^{-1})\right)\,.
\eaa
We can trace away the B region of the graph using the
following consequence of the commutation relations \eqref{c-comm}:
$\int d g_I \sigma(h_I g_I) \overline{\sigma(k_Ig_I)}
=\int d\gamma \delta(\gamma h_I k_I^{-1})$\,.
The result is
\begin{align*}
\rho^{(n)}_{red}
&=\int \prod_{i=1}^{n}
dh^{\va A_i}_I
dk^{\va A_i}_I  \sigma_{\va A_i}\!(h^{\va A_i}_I g^{\va A_i}_I )
\overline{\sigma_{\va A_i}\!(k^{\va A_i}_I g'^{\va A_i}_I )}\n\\
&\times\,
\delta(h^{\va v_i}_{1}(k^{\va v_i}_1)^{-1})
\prod_{J=2}^4 \delta(h^{\va v_i}_{J}(h^{t_{J^{\va v_i}}}_J)^{-1})
\delta(k^{\va v_i}_{J}(k^{t_{J^{\va v_i}}}_J)^{-1})
\,.
\end{align*}
The mixed nature of $\rho^{(n)}_{red}$ is encoded 
\emph{only}
in the relation $h^{\va A_i}_1=k^{\va A_i}_1$. 
This example shows a remarkable general property of these states:
\emph{the information about the combinatorial and geometric structure of the graph B is irretrievably
lost, as a consequence of \eqref{c-comm}}. 
This feature implements naturally the holographic features of null surfaces in classical gravity, and thus indirectly confirms the geometric interpretation of our GFT states. 
This happens even with no characterization of our states as a quantum horizon states, and it seems to follow directly
from the hypothesis of condensation, encapsulated in the operators \eqref{c-field}. Thus, it suggests that GFT condensates constitute a special class of 
holographic states.
\vspace{0.07cm}

{\bf Entropy.}---The computation of the entanglement entropy can be done in detail, as we are able to diagonalize the reduced density matrix. We work at fixed (large) number of vertices, which is compatible
with the semiclassicality conditions (semiclassicality requires anyway good peakednesss properties for the number operator, as this translates into good peakedness of extensive geometric observables). 
Using again \eqref{c-comm}, we see that the states
\bee\la{eigenstates}
\Psi^{(n)}_A
= \int \prod_{i=1}^{n}dg'^{\va A_i}_I df^{\va A_i}_I
\sigma_{\va A_i}(f^{\va A_i}_I g'^{{\va A_i}}_I )
\prod_{v,e} \delta(f_{v,e}f_{t_{e^{\va v}},e}^{-1})
\eee
are eigenstates of the horizon density matrix $\rho^{(n)}_{red}$.
Therefore, we can write the reduced density matrix of the horizon for a given number $n$ of boundary vertices as
\be\la{red-tot}
\rho_{red-tot}^{(n)}=\frac{1}{\mathcal N} \sum_{s=1}^{\mathcal N} \rho^{(n)}_{red}(\Gamma_s)\,,
\ee
where ${\mathcal N}$ is the total number of horizon graphs for a given number of vertices $n$,
obtained with the refinement operators, and $\rho^{(n)}_{red}(\Gamma_s)$ is the reduced density matrix 
for given graph. Orthogonality of the states for different graphs $\Gamma_s, \Gamma_{s'}$, which can be shown by direct computation, implies that the eigenvalue of $\rho^{(n)}_{red}(\Gamma_s)$ on a state $\Psi_{r_0}^{(n)}(\Gamma_{s'})$ is 1 if $s=s'$ and 0 if $s\neq s'$.
Hence, the diagonal form of the density matrix allows us to compute the von 
Neumann entropy of the horizon. In particular, it implies that 
\emph{the horizon entanglement entropy is the same as the Boltzmann entropy}, obtained by 
counting the boundary graphs, whose combinatorics, due to the condensate hypothesis, encode all the relevant microscopic DOF. 
For a state
\be
\Psi_{r_0}^{(n)}(\Gamma_s)=
\frac{\move_{\va b}^{N_{\va b}} \move_{\va w}^{N_{\va w}} }{N_{\va b}! N_{\va w}!}\ket{seed}
\ee
with $N_{\va b}+N_{\va w}=2n$, 
the total number of graphs 
with $n+1$ black and $n+1$ white vertices that can be constructed by acting with the refinement operators is given by
\be\la{nu}
 {\mathcal N}(n) = 
 \sum_{m=0}^{2n}\frac{(2n +1)!}{m! (2n-m)!}
 =2^{2n}
 (2n +1)\,.
 \ee
In the counting leading to the result \eqref{nu} we assumed indistinguishability of the vertices, consistently with the condensate hypothesis and the form of our states.
{
If we now include the degeneracy $\Delta(a)$ of the single-vertex Hilbert space, 
the number of \emph{states} to be counted is
$
\tilde{\mathcal N}(n,a)={\mathcal N}(n)\Delta(a)
$, where $a$ is the single-vertex area expectation value \eqref{a} for an outer boundary face of the horizon dual to a radial link (we suppress the subscripts $J,r,s$ in the following to lighten the notation).

A point requires attention. We are implicitly assuming
that the only structure that is left in the state is the horizon shell, while 
one would expect the computation of the full entanglement entropy to involve the reduced density matrix bulk part up to the horizon shell as well. Performing the same calculation above, one would expect to obtain the number
of graphs in the bulk times the single-vertex Hilbert space degeneracy associated to all the bulk shells. 
However, the construction of our states makes this extra counting not necessary. 
In fact, the refinement operators are applied on the whole state, and act in such a way that every new vertex on a shell is matched to a new vertex in a neighboring shell. Consequently,
the actions of the refinement operators on different shells, and hence the number of graphs to be
counted, are perfectly coordinated. The counting of the graphs on a single shell, then,
exhausts the number of states.  Moreover, for the action of the refinement operators (through \eqref{c-comm}) to be correctly defined on each part of every shell, the form of the condensate wave-function and its functional dependence on the different colors have to be the same for the whole graph. This implies that the degeneracy factor $\Delta(a)$ covers the dimension of the space of allowed wave-functions also when the whole bulk is included in the calculation. 

Therefore, {\it the holographic principle is not assumed in our analysis, but it follows from the condensate hypothesis and the features of our construction.}
We conclude that the Boltzmann entropy ($\log(\tilde{\mathcal N})$) is
\begin{equation}\la{SB}
S(n,a) =
2 n 
\log(2) +\log(2n+1) +\log(\Delta(a))\,.
\end{equation}

{
In \eqref{SB}, the central result of our analysis, we recognize an area law, as the first term is
an extensive quantity proportional to the total number of plaquettes composing the horizon, and thus, for given average area for a single-plaquette $a$,  to the total area $\mathcal{A} = a \, n$ (and the degeneracy factor $\Delta(a)$ only contributes a constant shift).
It should be stressed that the structure of the result holds for any spherically symmetric
state, as we have not yet discussed extra horizon conditions. This also implies that there is no reason, yet, to require matching with the Bekenstein--Hawking entropy, 
\ie requiring our states to give a specific value for $a$. 
{Notice that area laws for the entanglement entropy for any smooth closed codimension two surface emerge in various situations 
\cite{Solo}}. In this sense 
the commutation relations \eqref{c-comm} acquire a physical meaning, ensuring consistency between the quantum features of our GFT condensates and expected properties of classical smooth geometries, confirming their interpretation.

Let us emphasize a crucial point in order to fully appreciate the result \eqref{SB}. In our analysis we take into account both the single-vertex Hilbert space DOF, as well as combinatorial DOF encoded in the sum over all the graphs, both in the boundary and in the bulk.  When considering a non-perturbative state with possibly an infinite number of DOF, like in our analysis, it is far from obvious that the expected physical properties, from a perturbative analysis point of view, remain valid. In the case of the entanglement entropy of the reduced density matrix obtained when summing over all the boundary and bulk graphs, the scaling behavior with the area, even when valid for a given graph as shown in \cite{KMS, Donnelly}, is not an obvious physical property at all in a context of random geometries \cite{Eisert}. 
To proceed beyond this point one should use the equations of motion to determine $n,a,\Delta(a)$, not fixed by the defining properties of the condensate states alone.
Even without the exact dynamics, we can make
 significant progress by imposing horizon boundary conditions.
As pointed out above, we have two possibilities. 

Using the IH boundary condition would a priori introduce an extra dependence of the degeneracy $\Delta$ on the total horizon area $\AIH$, as this enters the resulting constraint on the vertex wave-function $\sigma$. The area law, then, is not guaranteed and one needs a detailed analysis of the space of constrained $\sigma$s. This would be a highly nontrivial task. 
We use instead a maximum entropy principle, and we
determine the values of $a, n, \Delta(a)$ for the most generic state compatible with a fixed macroscopic value of $\AIH$. 
Compatibly with the semiclassicality conditions stated above, we  consider condensate states such that $n$ is large and, consequently, $a$ is small. 
Introducing the area constraint, 
we look for extrema of
$\Sigma(n,a,\lambda) = S(n,a) + \lambda\left({{\mathcal A}_{\va H}} - 2 a n\right)$, when varying with respect to $a,n, \lambda$. Let us point out that, if $\Delta(a)$ was known explicitly, then the system of equations would fully determine the free parameters $a,n, \lambda$ as functions of $\AIH$ and the microscopic parameters of the theory. This not being the case, we use one of the equations to determine $\Delta(a)$, thus leaving the final result dependent on the Lagrange multiplier $\lambda$. More precisely, 
we obtain  $a=\log(2)/\lambda, \Delta= c_0 \exp{(\lambda \AIH)},$ where $c_0$ is an irrelevant integration constant. As
a result, the
entropy is
\be
S(\AIH, \lambda)\sim 2\lambda \AIH+ \log\left( \AIH/a\right) \la{entropy}
\,.
\ee

{\it We obtained the desired area law from first principles.}


From the entropy result \eqref{entropy} we recover the semiclassical entropy formula by setting the Lagrange multiplier $\lambda=1/8\ell_{\va P}^2$.  Within our working assumption about the compatibility of the classical dynamics with our hydrodynamical approximation of GFT, this last step can be interpreted as a thermodynamical consistency condition. 
More precisely, exploiting the continuum (and semiclassical) geometric interpretation of our states,  
the  value of $\lambda$ above yielding the 
factor of $1/4$ in the area law  is obtained from the compatibility with the thermodynamic relation
$
\beta \!=\! {\partial S}/{\partial E}
$, where $\beta$ is the horizon temperature and $E$ its energy, which implies convergence between macroscopic GR dynamics and effective equations of motion derived from the  GFT dynamics  (see \cite{Bianchi} for a microscopic derivation of the Unruh temperature for a local Rindler horizon).

Let us clarify an important aspect of this final result. The value of $\lambda$ yielding the correct semiclassical result implies $a=\log(2)8\ell_{\va P}^2$, which is also consistent with our semiclassicality condition of $a$ small, i.e. large $n$ limit.  The (average) area $a$ for a single-vertex can be computed for each specific choice of our microscopic GFT condensate states. The agreement with this precise value is then a constraint selecting those states, among those solving also the dynamics of the theory, which admit a good semiclassical interpretation. In this way, the (implicit) dependence  of $a$ on the Immirzi parameter does {\it not} imply that the Bekenstein--Hawking formula is recovered only for a specific choice of $\gamma$. On the contrary, the leading term in the semiclassical entropy result remains explicitly {\it independent on $\gamma$}. This is a striking consequence of the GFT formalism. More precisely, the availability of a number operator (a purely GFT observable), and the possibility to construct and control condensate states incorporating a large (possibly infinite) superposition of graphs, rather than simple area eigenstates, represent key improvements over similar calculations in canonical LQG. The standard LQG calculation (with its dependence on $\gamma$) would be recovered for very special condensate states which are eigenstates of the horizon area.

\vspace{0.07cm}

{\it Remarks}.---We notice that $\ell_{\va P}$, appearing in the entropy through  $\lambda$, is going to be a function of the microscopic parameters of the theory, i.e. its dynamical coupling constants. These, in turn, are subject to renormalization in going from the microscopic definition of the theory to the effective continuum (and semiclassical) regime. 
To determine the flow of such parameters 
 is an active direction of current research 
in GFT
\cite{Renormalization}.

Finally, let us point out that the coefficient in front of the logarithmic correction 
depends directly on the form of the refinement operators in the microscopic definition of our condensates, which dictates the counting of graphs. 
Moreover, if we consider a more general mixed density matrix
containing a mixture of states with different $n$, 
for semiclassical mixtures peaked around some value $n_0$
the dominant area law contribution is robust and independent on any detail of the mixture; on the other hand, the numerical coefficient of the log  term  changes due to its  combinatorial origin, but  is still of order unit and independent on $\gamma$. 

\begin{acknowledgments}
DP wishes to acknowledge the Templeton Foundation for the supporting Grant No. 51876. LS has been supported by the Templeton Foundation through the Grant No. PS-GRAV/1401.
\end{acknowledgments}

\end{document}